\documentclass[reprint,amsmath,amssymb,aps,twocolumn]{revtex4}

\usepackage{epsfig,amsmath}
\usepackage{graphicx}
\usepackage{dcolumn}
\usepackage{bm}
\usepackage{bbold}

\newcommand{\beq}{\begin{equation}}
\newcommand{\eeq}{\end{equation}}
\newcommand{\beqa}{\begin{eqnarray}}
\newcommand{\eeqa}{\end{eqnarray}}

\newcommand{\la}{\langle} 
\newcommand{\ra}{\rangle}

\begin{document}
\title {Sudden Freezing and Thawing of Entanglement Sharing in a Shrunken Volume}
\author{Yi Ding}
\email{yiding993@gmail.com}
\author{Songbo Xie}
\author{Joseph H. Eberly}
\affiliation{Center for Coherence and Quantum Optics, and Department of Physics and Astronomy, University of Rochester, Rochester NY 14627, USA}

\begin{abstract}

Within the one-excitation context of two identical two-level atoms interacting with a common cavity, we examine the dynamics of all bipartite one-to-other entanglements between each qubit and the remaining part of the whole system. We find a new non-analytic  ``sudden" dynamical behavior of entanglement. Specifically, the sum of the three one-to-other entanglements of the system can be suddenly frozen at its maximal value or can be suddenly thawed from this value in a periodic manner. We calculate the onset timing of sudden freezing and sudden thawing under several different initial conditions. The phenomenon of permanent freezing for entanglement is also found. We also identify a non-trivial upper limit for the sum of three individual entanglements, which exposes the concept of entanglement sharing in a shrunken ``volume''. Further analyses about freezing and thawing processes reveal quantitative and qualitative laws of entanglement sharing.
\end{abstract}

\maketitle

\section{Introduction}
Modern advances in cryptography and communication have come to rely heavily on practices based on nonclassical principles, specifically quantum principles. The scientific element commonly engaged for this is a form of correlation called quantum entanglement, which allows the transmission of messages that are immune to known deciphering strategies [1]. Recently, a different form of quantum correlation, called ``quantum discord" [2, 3], has been suggested as a replacement for entanglement for its startling ability to become ``frozen", namely to remain constant for a finite time, which has been studied both theoretically \cite{Datta, Mazzo, Mazi, Pinto, Cianciar} and experimentally \cite{Silva, Paula}. The following studies also report the possibility of freezing for entanglement in different scenarios\cite{AliRau, WuXu, quantum1, quantum2, quantum3}. All of these studies, either for discord or entanglement, have been concerned with freezing under the influence of one or another decoherence mechanism.

However, many conceptual questions still remain. For example, is decoherence necessary for freezing?  Whether freezing is a phenomenon reserved for mixed states? In the present work, we provide definitely negative answers to both of these open questions and shed light on a new recognition for the physical mechanism of freezing. We will demonstrate sudden freezing of entanglement in a lossless dynamical context. The demonstration also introduces entanglement sudden thawing in the same pure-state lossless context. We show that it is sufficient to focus on bipartite entanglements that arise in three-qubit interactions that are accessible via familiar cavity QED experimental arrangements. Our results demonstrate that the presence of decoherence is not needed to obtain entanglement freezing. Also, permanent freezing of entanglement in the three-party system is also possible. This can happen when the entire system is frozen to a ``cold'' state when no dynamical transmission is allowed.

Furthermore, we identify a non-trivial upper limit for the sum of three bipartite entanglements, which can be identified as a shrunken entanglement ``volume" to be shared by the three parties in the current three-qubit scenario. Analyses of the dynamical behaviors of freezing and thawing reveal quantitative and qualitative relations of the entanglement sharing properties implied by them. 

\section{Physical Background}
\noindent{\it Physical Model. }The model we are considering is similar to that in \cite{sharing}, where two identical atoms labelled by $1$ and $2$ with transition frequency $\omega_0$, are located in a common high-Q cavity (labelled by 0) with frequency $\omega$, and are coupled to the cavity with the same coupling constant $g$. The ground and excited states for the atoms are denoted by $|g\ra$ and $|e\ra$. We write the bare Hamiltonian in the usual way as a sum of atoms and cavity contributions ($\hbar=1$)%
\begin{equation}\label{hat}
H_\text{at} = \omega_0 T_{ee}^{(1)} +\omega_0 T_{ee}^{(2)},\text{ and }%
H_\text{cav} =\omega a^{\dag}a,
\end{equation}
where $T_{ee}^{(i)}$ is the usual excited state occupation number
operator for atom $i$, defined as $T_{ee}^{(i)} = | e\ra_i\la e |_i$, while $a$ and $a^{\dag}$ are the annihilation and creation operators of the
cavity mode. The usual interaction Hamiltonian between the atoms and the cavity
(under the rotating wave approximation) is
\begin{equation}\label{hint}
H_\text{int}=g\left(  a^{\dag}T_{ge}^{(1)}+T_{eg}^{(1)}a\right)  +g\left(  a^{\dag
}T_{ge}^{(2)}+T_{eg}^{(2)}a\right),
\end{equation}
where  $T_{ge}^{(i)}$ and $T_{eg}^{(i)}$ are the atomic transition
operators, given by $T_{ge}^{(i)} = \left\vert g\right\rangle_i\left\langle e\right\vert_i $ and $T_{eg}^{(i)}$ = $\left\vert e\right\rangle_i\left\langle g\right\vert_i$.

The operator for the total number of excitations of the system, defined by the operator $M=a^{\dagger}a+T_{ee}^{(1)}+T_{ee}^{(2)}$ commutes with the total Hamiltonian and is thus an integral of motion. The interaction Hamiltonian \eqref{hint} only causes effective transitions among the eigenstates of $M$ with eigenvalue $m$, namely, $|m-2,e,e\rangle$, $|m-1,e,g\rangle$, $|m-1,g,e\rangle$, $|m,g,g\rangle$, where the three available slots for the ket state represent the cavity, atom 1 and atom 2 respectively.
In the basis of these eigenstates, the total Hamiltonian of the system takes on a block structure along the main diagonal:
\begin{widetext}
\begin{equation}\label{htot}
H_{tot}=\left[
\begin{array}
[c]{ccccccccc}%
0 & 0 & 0 & 0 & \cdots & 0 & 0 & 0 & 0\\
0 & \omega_{0} & 0 & g & \cdots & 0 & 0 & 0 & 0\\
0 & 0 & \omega_{0} & g & \cdots & 0 & 0 & 0 & 0\\
0 & g & g & \omega & \cdots & 0 & 0 & 0 & 0\\
\vdots & \vdots & \vdots & \vdots & \ddots & \vdots & \vdots & \vdots &
\vdots\\
0 & 0 & 0 & 0 & \cdots & (m-2)\omega+2\omega_{0} & \sqrt{m-1}g & \sqrt{m-1}g & 0\\
0 & 0 & 0 & 0 & \cdots & \sqrt{m-1}g & (m-1)\omega+\omega_{0} & 0 & \sqrt{m}g\\
0 & 0 & 0 & 0 & \cdots & \sqrt{m-1}g & 0 & (m-1)\omega+\omega_{0} & \sqrt{m}g\\
0 & 0 & 0 & 0 & \cdots & 0 & \sqrt{m}g & \sqrt{m}g & m\omega
\end{array}
\right].%
\end{equation}
\end{widetext}
The first $1 \times 1$ block on the diagonal with the single number $0$ implies no interaction between the state $|0, g, g\rangle$ and the others. The second block is a $3\times3$ matrix that represents the case of one excitation in the system. The other blocks along the main diagonal are $4\times4$ matrices for all $m\geq2$, as is shown in Eq. (\ref{htot}).

Clearly, we can add arbitrarily many more atoms of the same transition frequency and coupling strength. If all atoms are initially in their ground state and if we allow for only one single photon initially in the cavity, the behavior of dynamics for the entire system will not change greatly.

The solutions for arbitrary initial excitations can be found in \cite{sharing}. Here, in order to display our main interests, we focus on the case of a single initial excitation, i.e. $m=1$. It is clear that the cavity can then only occupy $|0\rangle$ and $|1\rangle$ states, which can be considered as a third qubit itself. To solve for the 3-qubit state dynamics we denote the general form of the wave function for the system as
\begin{equation}\label{psi}
\left\vert \psi(t)  \right\rangle =a_{0}(t)\left\vert 1,g,g\right\rangle +a_{1}(t)\left\vert
0,e,g\right\rangle +a_{2}(t)\left\vert 0,g,e\right\rangle,
\end{equation}
where $a_{0}(t)$, $a_{1}(t)$ and $a_{2}(t)$ in
turn represent the probability amplitudes that the cavity, atom $1$ and atom $2$ are in the excited state. When substituting these into the Schr\"{o}dinger's equation, we obtain three first order differential equations,
\begin{equation}\label{1order}
    \begin{split}
\dot{\widetilde{a}}_1(t)  &  =-ig\widetilde{a}_{0}(t)e^{i\Delta t},\\
\dot{\widetilde{a}}_2(t)  &  =-ig\widetilde{a}_{0}(t)e^{i\Delta t},\\
\dot{\widetilde{a}}_0(t)  &  =-ig\left[  \widetilde{a}_{1}(t)+\widetilde{a}_{2}(t)\right]
e^{-i\Delta t},
\end{split}
\end{equation}
where $\widetilde{a}_{1}(t)$, $\widetilde{a}_{2}(t)$ and $\widetilde{a}_{0}(t)$ are the slowly varying amplitudes, defined as  $\widetilde{a}_{1}(t)=a_{1}(t)e^{i\omega_{0} t}$, $\widetilde{a}_{2}(t)=a_{2}(t)e^{i\omega_{0} t}$, $\widetilde{a}_{0}(t)=a_{0}(t)e^{i\omega t}$, and $\Delta$ is the detuning parameter defined as $\Delta
\equiv\omega_{0}-\omega.$ A simple solution of these equations, valid for an
arbitrary initial state is given by
\begin{align}\label{tilde}
\widetilde{a}_{1}(t)  &  =\widetilde{a}_{1}(0)-\alpha+\left\{  \alpha\cos(\frac{\Omega}{2}%
t)-i\beta\sin(\frac{\Omega}{2}t)\right\}  e^{i\frac{\Delta}{2}t}\nonumber,\\
\widetilde{a}_{2}(t)  &  =\widetilde{a}_{2}(0)-\alpha+\left\{  \alpha\cos(\frac{\Omega}{2}%
t)-i\beta\sin(\frac{\Omega}{2}t)\right\}  e^{i\frac{\Delta}{2}t}\nonumber,\\
\widetilde{a}_{0}(t)  &  =\left\{  \widetilde{a}_{0}(0)\cos(\frac{\Omega}{2}t)-i\gamma\sin
(\frac{\Omega}{2}t)\right\}  e^{-i\frac{\Delta}{2}t},
\end{align}
where $\Omega=\sqrt{\Delta^{2}+G^{2}}$ is a detuned Rabi oscillation frequency
with $G=\sqrt{8}g$. Also, $\widetilde{a}_{0}(0),$ $\widetilde{a}_{1}(0)$ and $\widetilde{a}_{2}(0)$ are the initial values
of the slowly varying amplitudes, with
\begin{equation}\label{alphabetagamma}
\begin{gathered}
\alpha=\frac{4g^{2}[\widetilde{a}_{1}(0)+\widetilde{a}_{2}(0)]}{\Omega^{2}
-\Delta^{2}},\\[1.5em]
\beta=\frac{4g^{2}\Delta\lbrack \widetilde{a}_{1}(0)+\widetilde{a}_{2}(0)]+2g\left(  \Omega^{2}%
-\Delta^{2}\right)  \widetilde{a}_{0}(0)}{\Omega\left(  \Omega^{2}-\Delta^{2}\right)  },\\[1.5em]
\gamma=\frac{2g[\widetilde{a}_{1}(0)+\widetilde{a}_{2}(0)]-\Delta \widetilde{a}_{0}(0)}{\Omega}.
\end{gathered}
\end{equation}

Throughout this paper, we consider only the resonant situation $\Delta=0$ for simplicity. This is enough to demonstrate the main results. In this way, the three original amplitudes $a_0(t)$, $a_1(t)$ and $a_2(t)$ differ from their slowly varying forms $\widetilde{a}_0(t)$, $\widetilde{a}_1(t)$ and $\widetilde{a}_2(t)$ only by a global phase. Thus, the slowly varying solutions can be changed just by the rapidly varying ones.\\

\noindent{\it Entanglement Measure. }In the three-qubit scenario, there exist various types of entanglement concerning different compositions of entangled parties, including tripartite entanglement, one-to-other bipartite entanglement and one-to-one bipartite entanglement within any two-qubit subsystems. We are here concerned with the dynamics of one-to-other bipartite entanglements, from the perspective of entanglement sharing. In what follows, we will adopt as our measure of entanglement the Schmidt weight  $K(t)$ \cite{KDef}, but in the normalized form, labelled as $Y(t)$, which is defined in the earlier work \cite{QAE} as
\begin{equation}\label{ydef}
Y=1-\sqrt{\frac{2}{K}-1},
\end{equation}
where%
\begin{equation}\label{kdef}
K=\frac{1}{\mu_{1}^{2}+\mu_{2}^{2}},
\end{equation}
with $\mu_{1}$ and $\mu_{2}$ being the
corresponding  eigenvalues of a Schmidt-bipartitioned reduced density matrix. $Y$ is monotonic with the concurrence $C$ \cite{Woott} via $Y = 1 - \sqrt{1 - C^2}$, and is a good entanglement measure.

With this definition in hand, we can express the three one-to-other bipartite entanglements in the forms
\begin{equation}\label{y0y1y2}
\begin{split}
Y_{0(12)}=2\min\left(  \left\vert a_{1}\right\vert ^{2}+\left\vert
a_{2}\right\vert ^{2},\left\vert a_{0}\right\vert ^{2}\right),\\
Y_{1(02)}=2\min\left(  \left\vert a_{0}\right\vert ^{2}+\left\vert
a_{2}\right\vert ^{2},\left\vert a_{1}\right\vert ^{2}\right),\\
Y_{2(01)}=2\min\left(  \left\vert a_{1}\right\vert ^{2}+\left\vert
a_{0}\right\vert ^{2},\left\vert a_{2}\right\vert ^{2}\right).
\end{split}
\end{equation}

\section{Entanglement Sharing in a Shrunken Volume $\mathcal{V}$}
The idea of entanglement sharing was established and developed in \cite{monogamy,qudits,sharing}. If a pair of parties A and B are maximally entangled, then neither A nor B can be entangled with a third party C. There exists a quantitative entanglement monogamy inequality marking the trade-off between the AB entanglement and the AC or BC entanglement. More specifically, the three one-to-one bipartite entanglements are restricted by the three one-to-other bipartite entanglements, namely
\begin{equation}
\begin{split}
    &C^2_{AB}+C^2_{AC}\leq C^2_{A(BC)},\\
    &C^2_{BA}+C^2_{BC}\leq C^2_{B(AC)},\\
    &C^2_{CA}+C^2_{CB}\leq C^2_{C(AB)}.
\end{split}
\end{equation}

Here, the measure of entanglement being used is the squared concurrence or 2-tangle. We have found that by using the normalized Schmidt weight measure, one can discover a fourth inequality where the three one-to-other bipartite entanglements \eqref{y0y1y2} are also restricted, which exposes a new kind of entanglement sharing.

Although the three one-to-other bipartite entanglements \eqref{y0y1y2} are collective properties of the whole three-party system, they have one-to-one correspondences with the three parties of the system. In that sense, one can also identify $Y_{i(jk)}$ as an individual property that is assigned to the party $i$, and denote it as $Y_i$. $Y_i$ is then interpreted as the entanglement owned by the party $i$. Individually, the maximal amount of entanglement that one party can have is 1:
\begin{equation}\label{singley}
    Y_i\leq1.
\end{equation}
A first estimate for the maximal entanglement that the three parties can have in total is 3:
\begin{equation}\label{ysdef}
Y_{S}:=Y_{0}+Y_{1}+Y_{2}\leq3.
\end{equation}
Intuitively thinking, the three parties are sharing entanglements from a large dimensionless entanglement ``volume" $\mathcal V=3$. Each party can take some entanglement, less than or equal to 1, from the volume. The best they can do is for all of them to take 1 and exhaust the whole volume.
However, this sharing of entanglement is firmly restricted by an inherent condition. The three squared amplitudes that three $Y_i$s depend on have only two degrees of freedom since they are fixed (see \eqref{y0y1y2}) by the normalization condition $|a_0|^2+|a_1|^2+|a_2|^2=1$. Together with the single initial excitation, this leads to the result that the total entanglement volume able to be shared by the three parties has shrunk to $\mathcal V=2$ from $\mathcal V=3$,
\begin{equation}
    Y_{0}+Y_{1}+Y_{2}\leq 2\leq3.
\end{equation}
The second inequality here signals the shrinkage of the available entanglement volume, while the first inequality is the ``fourth'' inequality that was mentioned above, which indicates the sharing of an entanglement volume $2$. An incomplete sharing is possible when $Y_{S}<2$, under which condition the volume is not fully exhausted by the three parties. 

The restrictions on the entanglement sharing for the three parties can be viewed from two directions. Firstly, the entanglements that each party can take from the volume are no longer independent, and thus the relation \eqref{singley} is not valid. A more compact form is 
\begin{equation}\label{newsingley}
    Y_i\leq Y_j+Y_k,
\end{equation}
or $Y_i\leq Y_S/2$. This relation is anticipated by the entanglement polygon inequality given in \cite{QAE}. Secondly, the content of the volume is restricted as described above. When two of the parties both take 1 from the total volume, there is nothing left for the third party. An example is the state $\left(|0,e,g\rangle+|0,g,e\rangle\right)/\sqrt{2}$, where entanglement volume is exhausted by the two atoms.

\section{Freezing and Thawing Dynamics of Entanglement}
In the following, we will mainly take two classes of initial conditions to illustrate the time evolution of $Y_{S}$: for the initial excitation to be either on the cavity or on the atoms.

The first class has the initial state:
\begin{equation}\label{ini1}\left\vert\psi(0)\right\rangle=\left\vert1,g,g\right\rangle.\end{equation}
After some algebraic simplification, we can express the sum of the three one-to-other bipartite entanglements in this case as%
\begin{equation}\label{ys1}
Y_{S}(t)=\left\{
\begin{array}{cc}
2,  & 2k+\frac{1}{2}\leq
Gt/\pi\leq2k+\frac{3}{2}\\[1em]
2-2\cos(Gt),  & \text{otherwise},
\end{array}
\right.%
\end{equation}
where $k=0,1,2,...$ is any non-negative integer.

\begin{figure}[t]
\centering
\includegraphics[width=8.6cm]{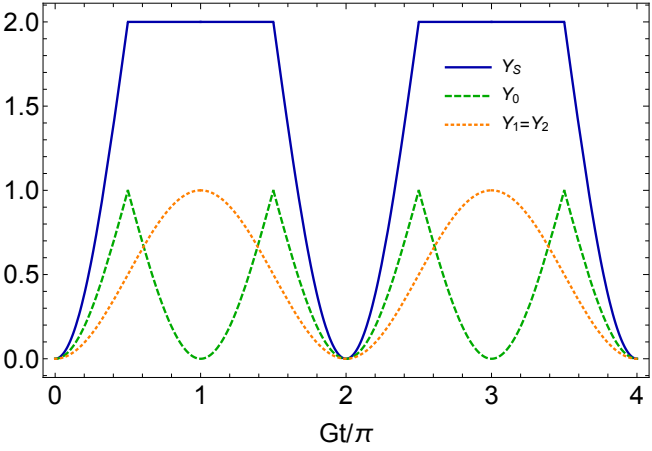}
\caption{For the initial state $|1,g,g\rangle$, the three bipartite entanglements $Y_i$ are shown as well as their sum, plotted as a function of the dimensionless time $Gt/\pi$. The sudden transitions to the flat top of $Y_S$ announce sudden freezings of entanglement. The solid line (blue in the color version) is for $Y_S$; the sharply-peaked dashed line (green in color) is for $Y_0$; and the smooth dotted line (orange in color) is for $Y_1$ and $Y_2$.}
\label{fig:fig1}
\end{figure}

In Fig. \ref{fig:fig1}, $Y_{S}(t)$ is plotted as a function of the dimensionless time $Gt/\pi$ as a solid line (blue in the color version) and is shown to increase in a monotonic fashion between $0$ and $Gt/\pi=0.5$ when it abruptly, with discontinuous slope, takes a steady value at the maximum $2$. It retains this value until $Gt/\pi=1.5$, when it starts to decay gradually to zero, also beginning in an abrupt fashion. We can call these dynamical behaviors entanglement sudden freezing (ESF) and entanglement sudden thawing (EST). 

The three individual entanglements are also plotted in Fig. \ref{fig:fig1}. It can be observed that the entanglement volume shared by the cavity $Y_{0}$ equals exactly the volume shared by the two atoms $Y_{1}+Y_{2}$ during thawing. Equivalently, we have $Y_{0} = Y_S/2$. Due to the symmetry of the two atoms in the initial condition, the two atoms always share the same amount from the volume in the whole process, namely $Y_S/4$. We can see from Fig. \ref{fig:fig1} that, in this thawing region, the total volume $\mathcal V=2$ is not completely shared by the three parties. On the contrary, in the freezing region, the volume is completely exhausted, and the sharing fractions of the three parties are strictly decided by the three squared amplitudes in ratios given by
\begin{equation}\label{ratio}
    Y_{0}:Y_{1}:Y_{2}=|a_0|^2:|a_1|^2:|a_2|^2.
\end{equation}\\
Since $Y_S$ is always 2 in freezing regions, the maximal value for an individual entanglement is 1. When $Y_{0}$ reaches the value 1 again, $Y_S$ thaws immediately, as can be seen at $Gt/\pi=1.5$.

The initial state for the second class is the partially entangled Bell state:
\begin{equation}\label{ini2}
|\psi(0)\ra=\cos\theta |0,e,g\ra + \sin\theta |0,g,e\ra,
\end{equation}
where the initial excitation is on the two atoms, and a parameter $\theta$ is introduced. After some long but straightforward calculations, we obtain the final expression of $Y_S(t)$, which is given as the sum of three distinct minimum functions:

\begin{widetext}
\begin{equation}\label{ys2}
Y_{S}(t)=2\left[
\begin{array}
{c}%
\min \Big\{ \Big(\frac{1+\sin(2\theta)}{2}\cos^{2}(\frac{Gt}{2})+\frac{1-\sin(2\theta)}%
{2} \Big),\ \ \Big(\frac{1+\sin(2\theta)}{2}\sin^{2}(\frac{Gt}{2}) \Big)\Big\}\\[1.5em]

+\min\Big\{ \Big(\frac{1+\sin(2\theta)}{4}\cos^{2}(\frac{Gt}{2}) + \frac{1+\sin(2\theta)}%
{2}\sin^{2}(\frac{Gt}{2}) + \frac{\cos(2\theta)}{2} \cos(\frac{Gt}{2}%
) + \frac{1-\sin(2\theta)}{4} \Big),\\[0.5em]
\Big( \frac{1+\sin(2\theta)}{4}\cos^{2}(\frac{Gt}{2})
-\frac{\cos(2\theta)}{2}%
\cos(\frac{Gt}{2})+\frac{1-\sin(2\theta)}{4} \Big) \Big\}\\[1.5em]

+\min\Big\{ \Big(\frac{1+\sin(2\theta)}{4} \cos^{2}(\frac{Gt}{2}) + \frac{1+\sin(2\theta)}%
{2} \sin^{2}(\frac{Gt}{2}) - \frac{\cos(2\theta)}{2} \cos(\frac{Gt}{2}%
) + \frac{1-\sin(2\theta)}{4} \Big),\\[0.5em]
\Big( \frac{1+\sin(2\theta)}{4}\cos^{2}(\frac{Gt}{2})+\frac{\cos(2\theta)}{2}%
\cos(\frac{Gt}{2})+\frac{1-\sin(2\theta)}{4} \Big) \Big\}
\end{array}
\right].
\end{equation}
\end{widetext}


Remarkably enough,  except in the cases of the perfectly symmetric ($\theta/\pi=0.25$) and  antisymmetric ($\theta/\pi=0.75$) Bell states, $Y_{S}(t)$ contains a new oscillation term $\cos(Gt/2)$ which suggests that the period with respect to the dimensionless time $Gt/\pi$ will be twice as long as that in Eq. \eqref{ys1}, namely $4$. However, as we will see, it is still $2$. The appearance of the new oscillation term will only effect the period of the system's quantum state \eqref{psi}, not the period of $Y_S$. The reason is mainly due to the symmetric situations of the two atoms, so that when the two atoms swap their situations, $Y_S$ doesn't change, but the dynamics are different.

\begin{figure}[t]
\centering
\includegraphics[width=8.6cm]{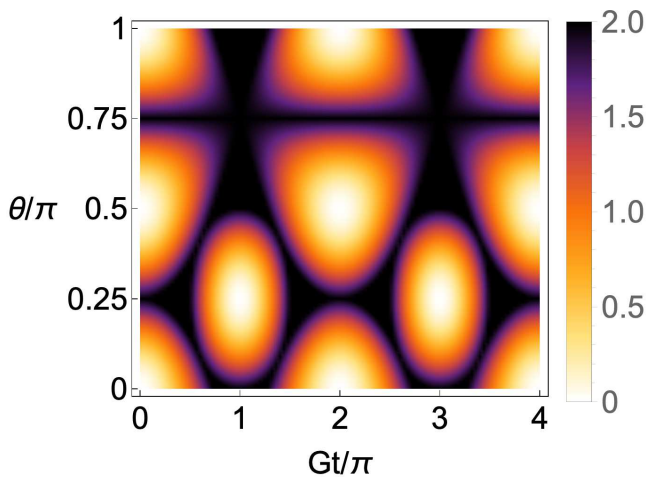}
\caption{The density plot of $Y_S(t)$ for the initial state \eqref{ini2}, as a function of both $Gt/\pi$ and the parameter $\theta/\pi$. The analytical expression is given by Eq. \eqref{ys2}. The connected dark regions in the figure (deep purple in the color version) locate freezing of entanglement, while the isolated white regions (red-yellow-white as colors) are for thawing of entanglement.}
\label{fig:fig2}%
\end{figure}

For the second class of initial conditions \eqref{ini2}, Fig. \ref{fig:fig2} shows the detailed behavior of $Y_S$ as a function of the dimensionless time $Gt/\pi$ and the parameter $\theta/\pi$. Freezing of entanglement $Y_S = 2$ is confined to the connected dark regions in the figure (deep purple in the color version) and the isolated white regions (red-yellow-white as colors) are for thawing of entanglement, with $Y_S < 2$. It is obvious that for all the values of $\theta$, the period of repeat with respect to $Gt/\pi$ is $2$, justifying the fact that the new extra term $\cos(Gt/2)$ doesn't double the period of $Y_S$. 

We note that the parameter $\theta$ can effectively control the freezing length of $Y_S$ within each period. At $\theta/\pi\approx0.4$, the freezing time is about to be minimized, when there is barely any freezing. On the contrary, at the specific value of $\theta/\pi=0.75$, $Y_S$ is permanently frozen and has a constant value 2. This is because the atoms are in the antisymmetrical Bell state, or in this case the ``cold'' state, for which the matrix elements of the interaction Hamiltonian are zero, meaning no dynamical evolution for the three-party system. From the viewpoint of entanglement sharing, this permanent freezing corresponds to the case that the total entanglement volume is completely exhausted by the two atoms.

The case $\theta/\pi=0.25$ is worth mentioning. It shows that $Y_{S}(t)$ is first frozen at the maximal value $2$, and then suddenly thaws at $Gt/\pi=0.5$. After a half cycle $Y_S$ is suddenly frozen at $2$ again. From the symmetry of the Hamiltonian as well as the symmetry of the initial state, we know that the two atoms evolve in an exactly identical way in this case. At $Gt/\pi=1$, the system's excitation is completely transferred to the cavity, and up to a global phase the state is found to be $|1,g,g\rangle$. Note that this is exactly the initial condition \eqref{ini1}. So the evolution picture for $Y_{S}(t)$ with the current initial condition can be obtained just by shifting the time axis of Fig. \ref{fig:fig1} by $1$. It is clear that the implicit laws of entanglement sharing behind the freezing and thawing processes are thus the same as that of the first class of initial states \eqref{ini1}.

\begin{figure}[t]
\centering
\includegraphics[width=8.6cm]{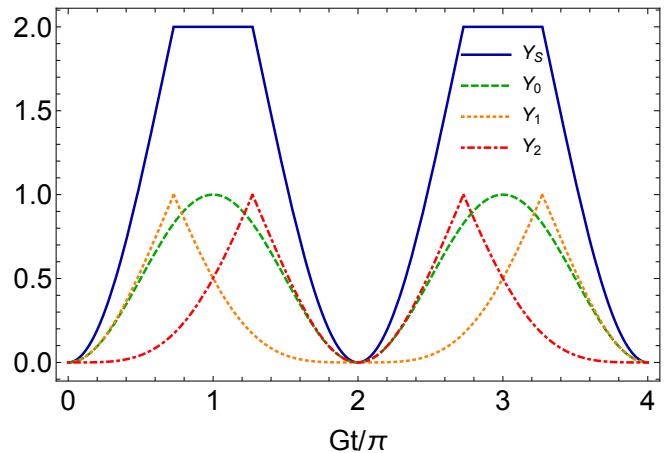}
\caption{For the initial state $|0,e,g\rangle$, the three bipartite entanglements $Y_i$ are shown as well as their sum, plotted as a function of the dimensionless time $Gt/\pi$. The freezing time can be seen to be shorter than that in Fig. \ref{fig:fig1}. The solid line (blue in the color version) is for $Y_S$; the dashed line (green in color) is for $Y_0$; the dotted (orange in color) line is for $Y_1$; and the dotdashed line (red in color) is for $Y_2$.}
\label{fig:fig3}%
\end{figure}

As a final example, we consider the special asymmetric case when the initial excitation is entirely on atom 1, namely $\theta/\pi=0$. The concrete dynamics of $Y_S$ as a function of the dimensionless time is displayed in Fig. \ref{fig:fig3}. Although the period doesn't change, one can see that the total freezing time within one period is shorter than that in Fig. \ref{fig:fig1}. During the first cycle of $Y_S$, the initial excitation in atom $1$ makes a complete transition to atom $2$ at $Gt/\pi=2$, and the excitation goes back to the atom $1$ at $Gt/\pi=4$ within the second cycle. The symmetry of the two atoms guarantees that the dynamics of $Y_S$ are identical for these two cycles. The complete expression for $Y_{S}(t)$ under this case is given by 
\begin{equation}
Y_{S}(t)=\left\{
\begin{array}
{cl}%
- \frac{1}{2}\cos\left(  Gt\right)  -2\cos\left(  \frac{Gt}{2}\right)
+ \frac{5}{2}, &
\tau_0\leq Gt/\pi\leq\tau_1\\[1em]
 
2, & 
\tau_1\leq Gt/\pi\leq\tau_2\\[1em]

- \frac{1}{2}\cos\left(  Gt\right)  + 2\cos\left(  \frac{Gt}{2} \right)
+ \frac{5}{2}, & \tau_2\leq Gt/\pi\leq\tau_3,
\end{array}
\right.  
\end{equation}
where $\tau_0=2k$, $\tau_1= 2k+ 2\arccos(\sqrt{2}-1)/\pi$, $\tau_2 = 2k+ 2\arccos(1-\sqrt{2})/\pi$, $\tau_3 = 2\left(  k+1\right)$ and $k$ is any non-negative integer.
In fact, from Eq. \eqref{tilde}, it is not difficult to find that the physical process for all values of $\theta/\pi$ (except for $0.25$ and $0.75$) is that the two atoms will swap their situations at the end of the first cycle ($Gt/\pi=2$), up to a global phase factor. As a result, $Y_{S}(t)$ must have the same evolution in these two cycles, as shown in the case of $\theta=0$. This is why the cycle of $Y_{S}(t)$ is still equal to $2$ despite containing a new oscillation term $\cos(Gt/2)$ in contrast to Eq. \eqref{ys1}.

The values for the three individual entanglements are also plotted as functions of the dimensionless time in Fig. \ref{fig:fig3}. It can be seen that the two atoms in turn share one half of $Y_S$ for each thawing region where three parties also don't exhaust the total volume. When one of them reaches the value 1, a new freezing or thawing begins. Within the freezing region, the entanglement volume $\mathcal{V}$ is completely exhausted by the three parties, and the continued ratio relation \eqref{ratio} for the distribution of the three parties' entanglements still holds.

\section{Summary}
In summary, we have presented sudden quantum freezing and thawing of photon-atom-atom entanglement in a lossless background for pure-state systems, and we have demonstrated that a dissipative environment is not necessary for freezing. This study provides an analytical understanding of dynamical effects of entanglement freezing and thawing for a quantum system consisting of three qubits, with one qubit representing the cavity. Our results suggest that the freezing duration of entanglement can be flexibly controllable in cavity QED experimental arrangements: from barely any freezing to finite time or to permanence. In addition, this study also provides an analytical understanding of the sharing dynamics of an entanglement ``volume'' $\mathcal{V}$, where both the total sharing volume and each sharing party are restricted in ways that were explained in the text. This may be conductive to design more rational protocols for implementing multiple quantum tasks in a multiparty system to avoid not only excessive waste of entanglement but also overloading tasks on any particular quantum units. The generalization of our results to an arbitrary $N$-qubit context is under way,  with results to be reported subsequently.\\

\section*{ACKNOWLEDGEMENTS}

We thank Prof. X.-F. Qian for several valuable discussions. Financial support was provided by National Science Foundation grants PHY-1501589 and PHY-1539859 (INSPIRE).

\end{document}